\documentclass[aps,showpacs,preprintnumbers,amsmath,amssymb]{revtex4}
\usepackage{dcolumn}
\usepackage[dvips]{epsfig}
\usepackage{graphicx}
\usepackage{amssymb}
\usepackage{color}
\usepackage{enumerate}
\usepackage{subfigure}

\begin{document}
\baselineskip=0.8 cm
\title{\bf Effect of system energy on quantum signatures of chaos in the two-photon Dicke model}

\author{Shangyun Wang$^{1}$\footnote{shangyun\underline{ }wang@163.com}, Songbai Chen$^{1,2}$\footnote{Corresponding author: csb3752@hunnu.edu.cn}, Jiliang
Jing$^{1,2}$ \footnote{jljing@hunnu.edu.cn}}

\affiliation{$ ^1$ Department of Physics, Key Laboratory of Low Dimensional Quantum Structures
and Quantum Control of Ministry of Education, and Synergetic Innovation Center for Quantum Effects and Applications,
Hunan Normal University, Changsha, Hunan 410081, People's Republic
of China\\
$ ^2$Center for Gravitation and Cosmology, College of Physical Science and Technology,
Yangzhou University, Yangzhou 225009, People's Republic
of China}

\begin{abstract}
\baselineskip=0.5 cm
\begin{center}
{\bf Abstract}
\end{center}

 We have studied entanglement entropy and Husimi $Q$ distribution as a tool to explore chaos in the quantum two-photon Dicke model. With the increase of the energy of system, the linear entanglement entropy of coherent state prepared in the classical chaotic and regular regions become more distinguishable, and the correspondence relationship between the distribution of time-averaged  entanglement entropy and the classical Poincar\'{e} section has been improved obviously. Moreover, Husimi $Q$ distribution for the initial states corresponded to the points in the chaotic region in the higher energy system disperses more quickly than that in the lower energy system. Our result imply that higher system energy has contributed to distinguish the chaotic and regular behavior in the quantum two-photon Dicke model.

\end{abstract}

 \maketitle
\newpage
\section{Introduction}

It is well known that classical chaos is a kind of very important and complex motion with the high sensitivity to initial conditions, which appears in the nonlinearly dynamical systems and can be usually detected by methods including the Poincar\'{e} surfaces of section and the Lyapunov characteristic
exponents. However, the chaos in the quantum mechanics becomes more intriguing and challenging since
there is no general quantum counterpart of classical phase space trajectories due to uncertainty principle \cite{chao01,chao02,chao03,chao04}. Moreover, the scalar product between two nearby states with different initial conditions remains a constant for all time rather than the exponential divergence because of the unitary time-evolution operator. Therefore, it is very important how to explore and explain signatures of chaos in quantum system, which could help us to understand further the quantum dynamics itself and the correspondence principle between  classical and quantum mechanics.

With the Random Matrix Theory, Wigner \textit{et al} \cite{Wigner1,Wigner2} analysed statistical properties at different energy levels in a quantum chaos state, and found that the distribution of spacings between adjacent energy levels for quantum chaos states obeys Wigner distributions governed by the Gaussian ensemble of matrices rather than usual Poisson distribution. Moreover, the study of entanglement entropy indicates that in general the chaotic systems tend to own larger entanglement entropy than the regular systems \cite{en1,en2,en4,en5,t14}. However, the correspondence relationship between entanglement entropy and chaos is not always hold since there exist certain cases in which the entanglement entropy for the initial state prepared in the regular region is higher than that of in chaotic region \cite{exch1,exch2,exch3}. The intrinsic physics is still uncertain and needs to be further investigated.
Recently, Ruebeck \textit{et al} \cite{exch3} studied the entangling quantum kicked top  and divided the infinite-time-averaged entanglement entropy $S_Q$ into two parts: $I_Q$ and $R_Q$, which come respectively from the ``diagonal"  and ``off-diagonal" matrix elements of the angular momentum operators obtained by
the Floquet eigenstates of the system. They found that $I_Q$ and $S_Q$ were correlated with a quantity $I_c$, which is not equivalent to classical chaos. In the quantum kicked top model, Piga \cite{enchao1}\textit{et al} also found that with increasing the number of qubits the entanglement entropy of the initial states in the classical regular and chaotic regions become more distinguishable, which leads to a more clear correspondence between the entanglement entropy and the classical features of the phase space. Moreover, they also found that there exists certain similar behaviors between the low entanglement entropy tori and Kolmogorov-Arnol'd-Noser (KAM) tori, which implies that entanglement could play an
important role in quantum KAM theory. The efforts have been devoted to developing a quantum analogue of KAM theory in Refs.\cite{KAM1,KAM2,KAM3,KAM4}. Other quantum sources,  such as, spin squeezing \cite{t1}, quantum discord \cite{t2}, out-of-time ordered correlator \cite{t3,t4,t5} have been applied as signatures to explore chaos in quantum systems.

In this paper, we will focus on  the two-photon Dicke model where
$N$ identical two-level atoms couple to a bosonic mode through two-photon interaction. Such a kind of  two-photon interaction have been commonly applied to describe the second-order process in physical devices including quantum dots \cite{qdot1,qdot2}, trapped ions and Rydberg atoms in microwave superconducting cavities \cite{trap1,trap2} . Comparing with the standard Dicke model \cite{t6}, the presence of two-photon interaction results in some new properties appeared in this quantum system. For example, the discrete
system spectrum collapses into a continuous band for a specific
value of the coupling strength \cite{coll1,coll2,coll3}. In the transition
from the strong to the ultrastrong coupling  regime, a continuous symmetry  breaks down into a four-folded discrete symmetry described by a
generalized-parity operator \cite{coll4}. Moreover, a super-radiant phase transition \cite{t9} also occurs in the two-photon Dicke model due to coherent radiations of the atoms. The behavior of finite-size scaling functions \cite{t10} in two-photon Dicke model indicates that the super-radiant phase transition has the same scaling features as in the standard Dicke model. Since two-photon coupling is a nonlinear interaction, chaos phenomenon would appear in such dynamical systems. However, quantum signatures of chaos and the correspondence between entanglement and classical chaos are still open in the two-photon Dicke model. Moreover, besides increasing the particle number $N$ as in the quantum kicked top model, it is natural whether there are other way to improve the correspondence between chaos and entanglement entropy. This is very interesting and is necessary to be further investigated in a quantum system.  The main motivation in this paper is to study entanglement entropy and Husimi $Q$ distribution and probe further the relationship between entanglement and classical chaos. We find that with higher system energy the values of linear entanglement entropy of the points between in chaotic and regular regions become more distinguishable for the two-photon Dicke model. Meanwhile, the higher system energy improves obviously the correspondence relationship between the distribution of time-averaged  entanglement entropy and the classical Poincar\'{e} section in this model.

The paper is organized as follows. In Sec.II, we introduces briefly the two-photon Dicke model and its properties. In Sec.III, we study the effects of system energy on correspondence between the distribution of time-averaged  entanglement entropy and the classical Poincar\'{e} section in the two-photon Dicke model. In Sec.IV,  we analyze the Husimi $Q$ distribution and further probe the effects of system energy on quantum signatures of chaos in this model. Finally, we present results and a brief summary.

\section{THE TWO-PHOTON DICKE MODEL}

Let us now briefly introduce the two-photon Dicke model in which $N$ two-level identical atoms interact with a single bosonic mode through a two-photon interaction. The system Hamiltonian can be expressed as \cite{t6}
\begin{eqnarray}
\hat{H}&=&\omega\hat{a}^{\dagger}\hat{a}+\omega_{0}\hat{J}_{z}+
\frac{g}{N}(\hat{J}_{+}+\hat{J}_{-})(\hat{a}^{2}+\hat{a}^{\dagger2}),\label{Hami1}
\end{eqnarray}
where $\hat{a}$ and $\hat{a}^\dagger$, respectively, are the annihilation and creation operators of the single-mode cavity with frequency $\omega$. Here, $\omega_{0}$ is the atomic transition frequency and $g$ is
the collective coupling strength of the two-photon interaction. $\hat{J}_{z}=\sum_{n=1}^{N}\hat{\sigma}_{z}^{(i)}/2$ is
the two-level atomic inversion operator,
$\hat{J}_{+} = \sum_{n=1}^{N}\hat{\sigma}_{+}^{(i)}/2$ and $\hat{J}_{-} = \sum_{n=1}^{N}\hat{\sigma}_{-}^{(i)}/2$ are the collective atomic raising and lowing operators. The operators $\hat{J}_{z}$, $\hat{J}_{+}$ and $\hat{J}_{-} $ form the $SU$(2) Lie algebra, which obeys the commutation relations
\begin{eqnarray}
[\hat{J}_{+},\hat{J}_{-}]=2\hat{J}_{z},\;\;\;\;\;\;\;\;[\hat{J}_{z},\hat{J}_{\pm}]=\pm\hat{J}_{\pm}.
\end{eqnarray}
Comparing with the usual standard Dicke model, the Hamiltonian in the two-photon Dicke model owns a generalized $Z_{4}$ parity operator $\prod=(-1)^{N}\otimes_{n=1}^{N}\hat{\sigma}_{z}^{n}e^{i\pi\hat{a}^{\dagger}\hat{a}/2}$ with four eigenvalues ( i.e., $\pm1$ and $\pm i$ )  rather than the $Z_{2}$ parity in the standard Dicke model \cite{t11,t12}.

\begin{figure}[ht]
\includegraphics[width=13cm]{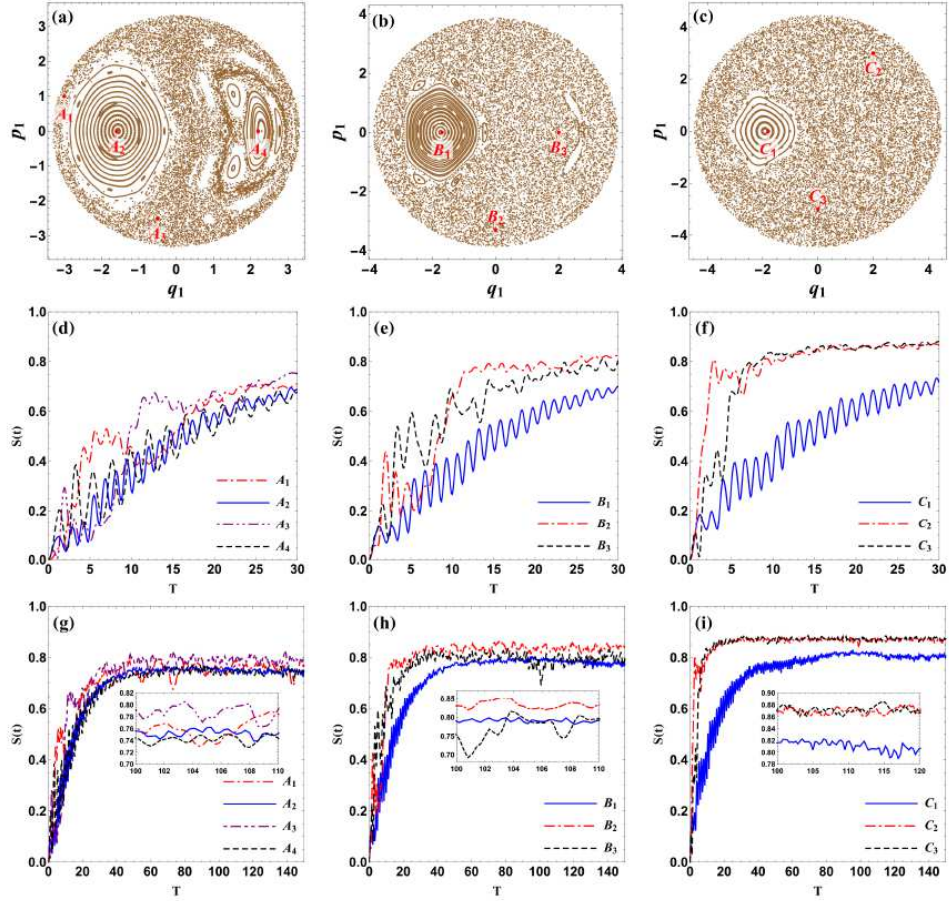}
\caption{(a)-(c) correspond to the classical phase spaces for the two-photon Dicke model with the system energy $E=1$, $5$, $10$, respectively. The points $A_1$, $A_3$, $B_2$, $B_3$, $C_2$ and $C_3$ are in classical chaotic regions, but the points $A_2$, $A_4$, $B_1$ and $C_1$ are in classical  regular regions.
(d)-(i) present the evolution of linear entanglement entropy $S(t)$ with time $t$ for these points, respectively. A comparison of (d)-(f) or (h)-(i)
shows that as one increases the energy of system the linear entanglement entropies of the points between in these chaotic and regular regions become more distinguishable. Here, we set $\omega=\omega_{0}/2=1$, $j=5$ and $g=0.3$.}\label{f1}
\end{figure}

In order to study the relationship between entanglement and classical chaos in the two-photon-Dicke model, we here take the initial states to be coherent states since they correspond to the minimum uncertainty wave packets centered in the classical phase space. As in Refs.\cite{en2,t11,t12,sd1}, we choose the initial quantum states as
\begin{eqnarray}
|\psi(0)\rangle&=&|\tau\rangle\otimes|\beta\rangle\equiv|\tau\beta\rangle\ ,
\end{eqnarray}
where $|\tau \rangle$  and  $|\beta \rangle$ are the atomic and bosonic coherent states,respectively. The coherent states $|\tau \rangle$  and  $|\beta \rangle$ own the forms
\begin{eqnarray}
|\tau\rangle&=&(1-\tau\tau^{*})^{-j}e^{\tau\hat{J}_{+}}|J,-J\rangle,\nonumber\\
|\beta\rangle&=&e^{-\beta\beta^{*}/2}e^{\beta\hat{a}^{\dagger}}|0\rangle,
\end{eqnarray}
with
\begin{eqnarray}
\tau&=&\frac{q_{1}+ip_{1}}{\sqrt{4j-q_{1}^{2}-p_{1}^{2}}},\nonumber\\
\beta&=&\frac{1}{\sqrt{2}}(q_{2}+ip_{2}).
\end{eqnarray}
Here $|J,-J\rangle$ denotes the state with spin $J$ and $\hat{J}_{z}=-J$, and $|0\rangle$ is the bosonic field ground state. The quantity $j$ is set to $j=N/2$ and the variables $q_{1},p_{1},q_{2},p_{2}$  describe the phase space of the system.
The indices $1$ and $2$ denote the atomic and field subsystem, respectively. With the standard procedure \cite{sd1}, one can obtain the corresponding classical Hamiltonian for the two-photon Dicke model (\ref{Hami1})
\begin{eqnarray}
H_{cl}\equiv\langle\tau\beta|\hat{H}|\tau\beta\rangle
=\frac{\omega_{0}}{2}(q_{1}^{2}+p_{1}^{2}-2j)+\frac{\omega}{2}(q_{2}^{2}
+p_{2}^{2})+\frac{q_{1}\sqrt{4j-q_{1}^{2}-p_{1}^{2}}(q_{2}^{2}-p_{2}^{2})g}{2j}.\label{Hcl}
\end{eqnarray}
And then, it is easy to get the canonical equations of Hamilton
\begin{eqnarray}
\dot{q}_{1}&=&\omega_{0}p_{1}+\frac{gq_{1}p_{1}(p_{2}^{2}-q_{2}^{2})}{2j\sqrt{4j-q_{1}^{2}-p_{1}^{2}}},
\nonumber\\
\dot{q}_{2}&=&\omega p_{2}-\frac{gq_{1}p_{2}\sqrt{4j-q_{1}^{2}-p_{1}^{2}}}{j},\nonumber\\
\dot{p}_{1}&=&-\omega_{0}q_1+
\frac{g(4j-2q_{1}^{2}-p_{1}^{2})(p_{2}^{2}-q_{2}^{2})}{2j\sqrt{4j-q_{1}^{2}-p_{1}^{2}}},\nonumber\\
\dot{p}_{2}&=&-\omega q_{2}-\frac{gq_{1}\sqrt{4j-q_{1}^{2}-p_{1}^{2}}}{j}.\label{motEq}
\end{eqnarray}
For the two-photon Dicke model, there exists a spectral collapse at $g_{collapse}=\frac{\omega}{2}$, which yields that the energy levels of the system collapse into a continuum as $g\geq\frac{\omega}{2}$ and then the ground state of Hamiltonian (\ref{Hami1}) is no longer defined in this regime. Therefore, we will focus on the strong-coupling regime, where $g$ is smaller, but comparable to $\omega/2$. For the sake of simplicity, we limit our consider on the resonant case $\omega_0=2\omega$ in which the transition between two energy levels occurs only if the atom absorbs (or emits) two photons. The presence of the two-photon interaction yields that the equation of motion (\ref{motEq}) in the classical correspondence are not be variable-separable, which means that the corresponding motion could be chaotic.
In Fig. \ref{f1}(a)-(c), we present the Poincar\'{e} section for certain parameters and initial values, which show that there exist chaos for system described by the classical Hamiltonian (\ref{Hcl}) corresponding to the two-photon Dicke model (\ref{Hami1}).

\section{Effects of Energy of system on quantum signatures of chaos in the two-photon Dicke model}

In this section, we will investigate how the system energy improve quantum signatures of chaos in the two-photon Dicke model by the linear entanglement entropy, which is a common tool to explore chaos in quantum systems including Dicke model \cite{en2} and Kicked top model \cite{t14}.
The linear entanglement entropy is defined as
\begin{eqnarray}
S(t)&=&1-Tr_{1}\rho_{1}(t)^{2},
\end{eqnarray}
with the reduced-density matrix
\begin{eqnarray}
\rho_{1}(t)=Tr_{2}|\psi(t)\rangle\langle\psi(t)|,
\end{eqnarray}
where $Tr_i$ is a trace over the $i$th subsystem ($i=1,2$), and the vector $|\psi(t)\rangle$ is the quantum state of the full system evolved in time under the action of Hamiltonian (1). The quantity $S(t)$ describes the degree of purity of the subsystems and the degree of decoherence.
In Fig.\ref{f1} (d)-(i), we show that the evolution of linear entanglement entropy with time $t$ for the different initial states which correspond to different points in the classical phase spaces. The points $A_1$, $A_3$, $B_2$, $B_3$, $C_2$ and $C_3$ are in classical chaotic region, but the points $A_2$, $A_4$, $B_1$ and $C_1$ are in classical  regular region. As the energy of system $E=1$, from Fig.\ref{f1}(d) and (g), it seems that the linear entanglement entropy increases more rapidly for the initial states corresponded to these points in the classical chaotic region. However, the values of linear entanglement entropy for the points $A_1$--$A_4$ are very close so that they are overlapped at sometime, which means that it is actually difficult to distinguish classical chaotic and regular behaviors by using of the linear entanglement entropy in this case. With the increase of the energy of system $E$, one can obtain that the linear entanglement entropies of the points in the chaotic and regular regions become more distinguishable. As $E=10$, from Fig.\ref{f1} (f) and (i), we find that with the time the linear entanglement entropy tends to different limit values for different initial states. The limit value of linear entanglement entropy for the states corresponded to the points ($C_2$, $C_3$) in chaotic regions is much higher than that of the points ($C_1$) in regular regions. This means that higher energy of system can enhance the availability of entanglement entropy as a tool to explore quantum chaos in the two-photon Dicke model.
\begin{figure}
\includegraphics[width=11cm]{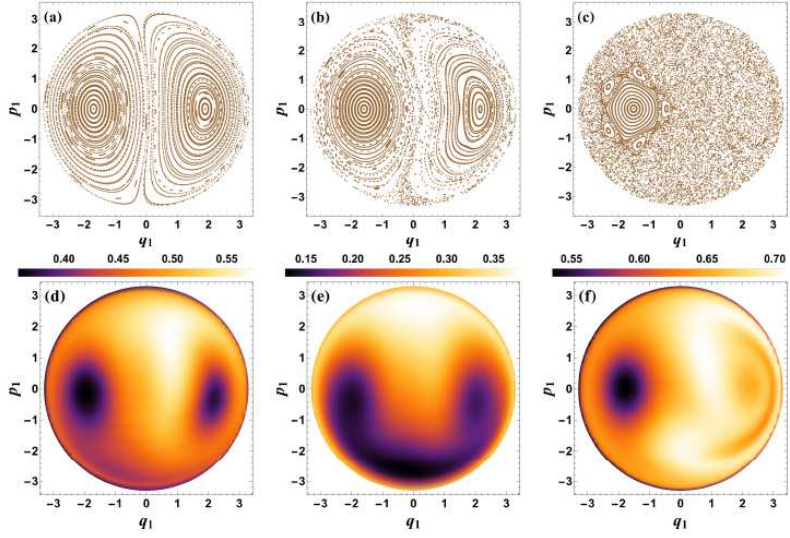}
\caption{(a)-(c) correspond to the classical phase spaces for the two-photon Dicke model with the coupling parameter $g=0.1$, $0.25$, $0.4$, respectively. (d)-(f) denote the corresponding time-averaged entanglement entropy distribution for the two-photon Dicke model with the coupling parameter $g=0.1$, $0.25$, $0.4$, respectively. Here, we set $\omega=\omega_{0}/2=1$, $j=5$ and $E=1$.}\label{f201}
\end{figure}

Now, we adopt the time-averaged entanglement entropy to investigation the correlation between classical
dynamics and quantum entanglement the two-photon Dicke model for different system energies. The time-averaged entanglement entropy is defined by
\begin{eqnarray}
S_{m}&=&\frac{1}{T}\int_{0}^{T}S(t)dt,
\end{eqnarray}
where $T$ is the total time of evolution.
\begin{figure}
\includegraphics[width=11cm]{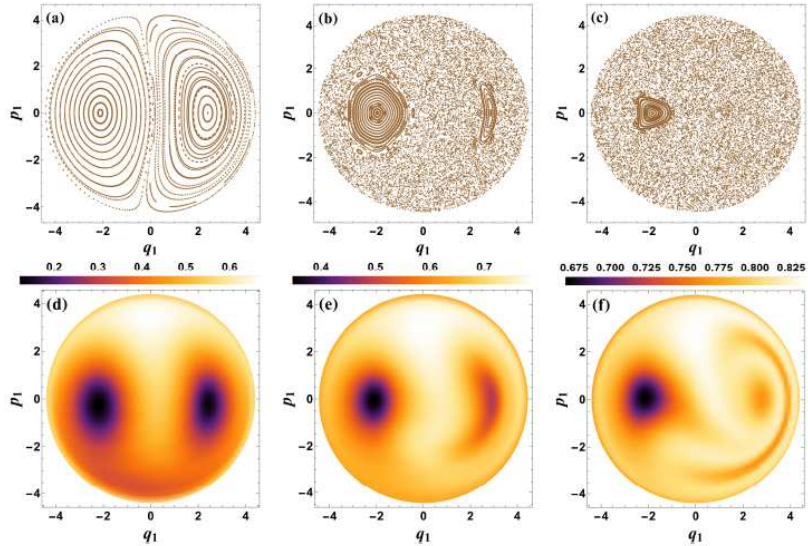}
\caption{(a)-(c) correspond to the classical phase spaces for the two-photon Dicke model with the coupling parameter $g=0.1$, $0.25$, $0.4$, respectively. (d)-(f) denote the corresponding time-averaged entanglement entropy distribution for the two-photon Dicke model with the coupling parameter $g=0.1$, $0.25$, $0.4$, respectively. Here, we set $\omega=\omega_{0}/2=1$, $j=5$ and $E=10$.}\label{f22}
\end{figure}
In Fig.\ref{f201} (a)-(c), we present the classical phase space for the two-photon Dicke model with the system energy $E=1$ for fixed coupling parameter $g=0.1$, $0.25$ and $0.4$, respectively. The corresponding time-averaged entanglement entropy $S_{m}$ with $T=30$ are plotted in Fig. \ref{f201} (d)-(f) for the initial states related to the points in the whole classical phase space. In Fig.\ref{f201}, comparing  figure (a) (or (c)) with figure (d) (or (f)),
it seems that there exist a correspondence between the classical phase space and the distribution of time-averaged  entanglement entropy, i.e., the initial state located in the chaotic region in classical phase space owns the high time-averaged  entanglement entropy and  the initial state in the regular region has the lower entanglement entropy. However, we also note that the time-averaged  entanglement entropy for certain points lied in the regular region is higher than that of in chaotic region. Actually,
this behavior of the time-averaged  entanglement entropy in the few-particle regime is also found in the quantum kicked top model \cite{Ghose2019}. It is shown that in the quantum kicked top system the semi-classical limit is approached by increasing the particle number $N$ and then the correspondence between the classical phase space and the distribution of time-averaged  entanglement entropy is improved. For the two-photon Dicke model, from Fig. \ref{f201}(b) and \ref{f201}(e), we can find directly that the correspondence relationship between the classical phase space and the time-averaged  entanglement entropy does not  hold in the few-particle case as the system energy is set to $E=1$. In Fig. \ref{f22}, we present the classical phase space and the time-averaged  entanglement entropy for the two-photon Dicke model with the same parameters as in Fig. \ref{f201} except the system energy $E=10$. Comparing  Figs. \ref{f201} and \ref{f22}, it is obvious that in the few-particle case ($N=10$) higher system energy  improves the correspondence between the classical phase space and the distribution of time-averaged entanglement entropy, which could be attributed to that at high system energy a quantum state would expect thermalization to a ``classical" high temperature state.

\begin{figure}
\includegraphics[width=14cm]{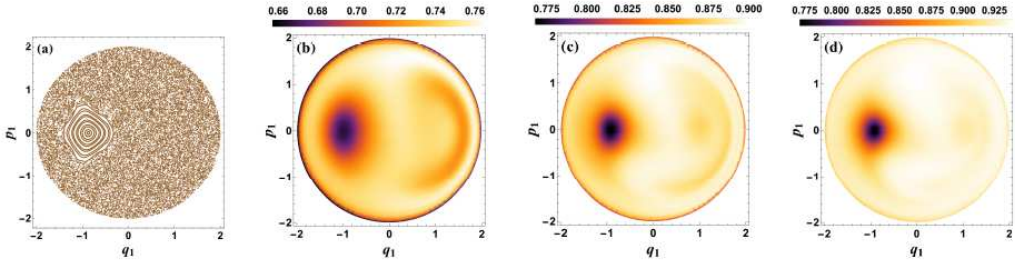}
\caption{The correspondence between classical dynamics and quantum entanglement entropy for the two-photon Dicke model with the fixed ratio $E/N=1.1$. (a) is the classical Poincar\'{e} section, and (b)-(d) denote the corresponding time-averaged entanglement entropy distribution  for the two-photon Dicke model with $E=5.5, N=5$; $E=16.5, N=15$ and $E=33, N=30$, respectively. Here we rescale $q1\rightarrow q1/\sqrt{j}$, $p1\rightarrow p1/\sqrt{j}$ and set the coupling parameter $g=0.3$ and $\omega=\omega_{0}/2=1$.}\label{f3}
\end{figure}

It is well known that in Kicked top model the structure of the classical phase space does not depend on the number of particle $N$ \cite{t14} and then it is  convenient to compare the correlation between classical dynamics and quantum entanglement for the systems with different $N$. However, in the two-photon Dicke model, one can find from Fig.\ref{f1} that the system energy affects the structure of the classical phase space. Actually, the structure of the classical phase space in the Dicke model does not change if the ratio $E/N$ is fixed. In Fig.\ref{f3}, we present the correspondence between classical dynamics and quantum entanglement for the two-photon Dicke model with the fixed ratio $E/N=1.1$.
We find that the correspondence is improved by increasing the energy $E$ and the particle number $N$ of system at the same time. However, it is unclear which factor (the energy $E$ or the particle number $N$) is responsible for this improvement. Thus, in order to probe effects of the system energy $E$ on the correspondence between classical dynamics and quantum entanglement entropy, as in the previous discussion, we have to change the system energy $E$ and fix the particle number $N$ in the two-photon Dicke model although it will change the structure of the classical phase space. Actually, we can compare the similarity degree between the classical Poincar\'{e} section and the distribution of quantum entanglement in the cases with different system energy and then probe further their correspondence.
\begin{figure}[ht]
\includegraphics[width=11cm]{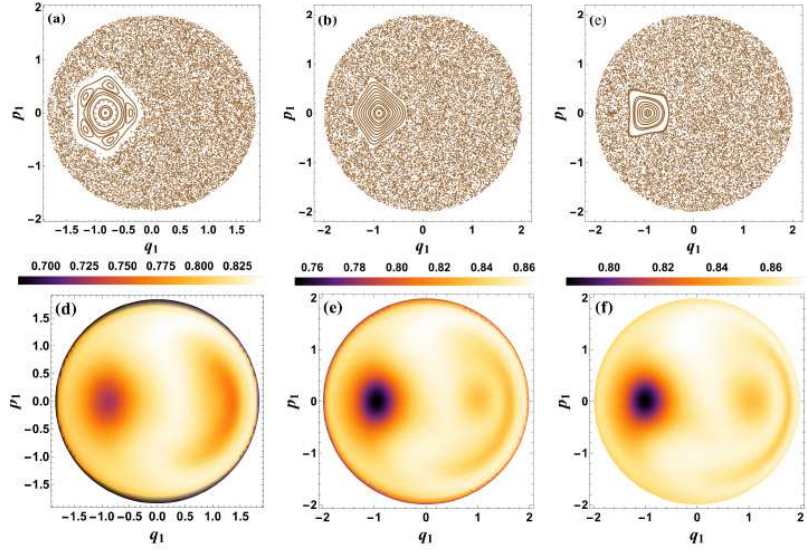}
\caption{The correspondence between classical dynamics and quantum entanglement entropy for the two-photon Dicke model with the fixed particle number $N=10$ and coupling constant $g=0.3$. (a)-(c) correspond to the classical phase spaces for the two-photon Dicke model with the system energy $E=7$, $12$, $17$, respectively. (d)-(f) denote the time-averaged entanglement entropy for the two-photon Dicke model with the system energy $E=7$, $12$, $17$, respectively. Here we set $\omega=\omega_{0}/2=1$. }\label{f3s1}
\end{figure}
\begin{figure}
\includegraphics[width=11cm]{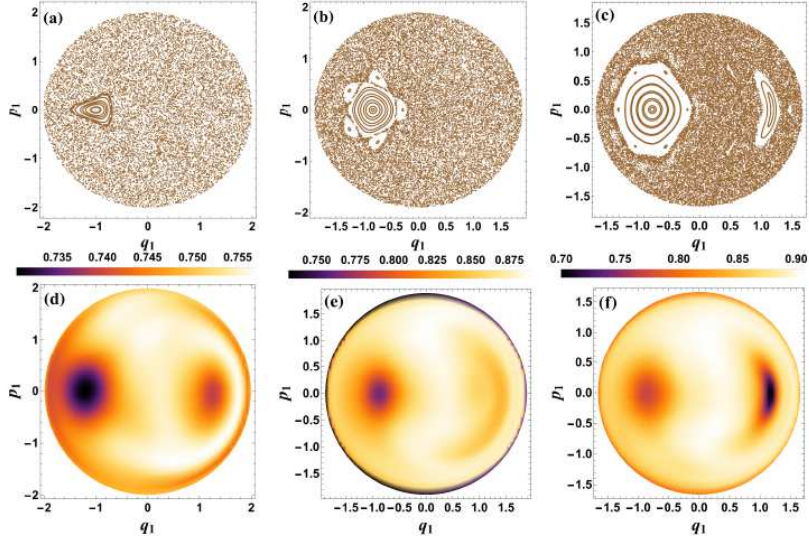}
\caption{The correspondence between classical dynamics and quantum entanglement entropy for the two-photon Dicke model with the fixed system energy $E=12$ and coupling constant $g=0.3$. (a)-(c) correspond to the classical phase spaces for the two-photon Dicke model with the particle number $N=4$, $15$, $30$, respectively. (d)-(f) denotes the time-averaged entanglement entropy for the two-photon Dicke model with the particle number $N=4$, $15$, $30$, respectively. Here we set $\omega=\omega_{0}/2=1$.}\label{f3s2}
\end{figure}
In Fig.\ref{f3s1}, we present the correspondence between classical dynamics and quantum entanglement entropy for the two-photon Dicke model with the fixed particle number $N=10$ and the coupling constant $g=0.3$. It is obvious that the increase of $E$ enhances the similarity degree between the classical Poincar\'{e} section and the distribution of quantum entanglement entropy. Moreover, we find that with the increase of $E$, the time-averaged entanglement entropy for the initial state near the boundary in the classical phase space, which is in the chaotic region,  becomes gradually higher than that for the initial state in the regular region. In Fig.\ref{f3s2}, we find that the increase of the particle number $N$ also improves the correspondence between classical dynamics and the time-averaged entanglement entropy, which is similar to those obtained in Kicked top model \cite{t14}.
\begin{figure}[ht]
\includegraphics[width=15cm]{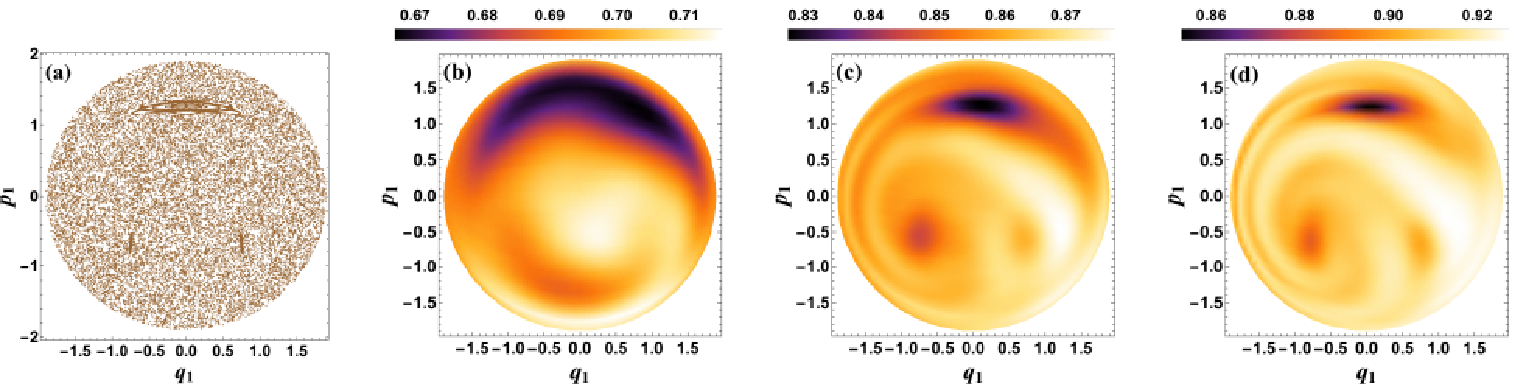}
\caption{The correspondence between classical dynamics and quantum entanglement entropy for the one-photon Dicke model with the fixed ratio $E/N=0.4$. (a) is the classical Poincar\'{e} section, and (b)-(d) denote the time-averaged entanglement entropy for the one-photon Dicke model in the cases with $E=2, N=5$; $E=6, N=15$ and $E=12, N=30$, respectively. Here we rescale $q1\rightarrow q1/\sqrt{j}$, $p1\rightarrow p1/\sqrt{j}$ and set the coupling parameter $g=0.5$ and $\omega=\omega_{0}=1$.}\label{fd3}
\end{figure}
\begin{figure}[ht]
\includegraphics[width=11cm]{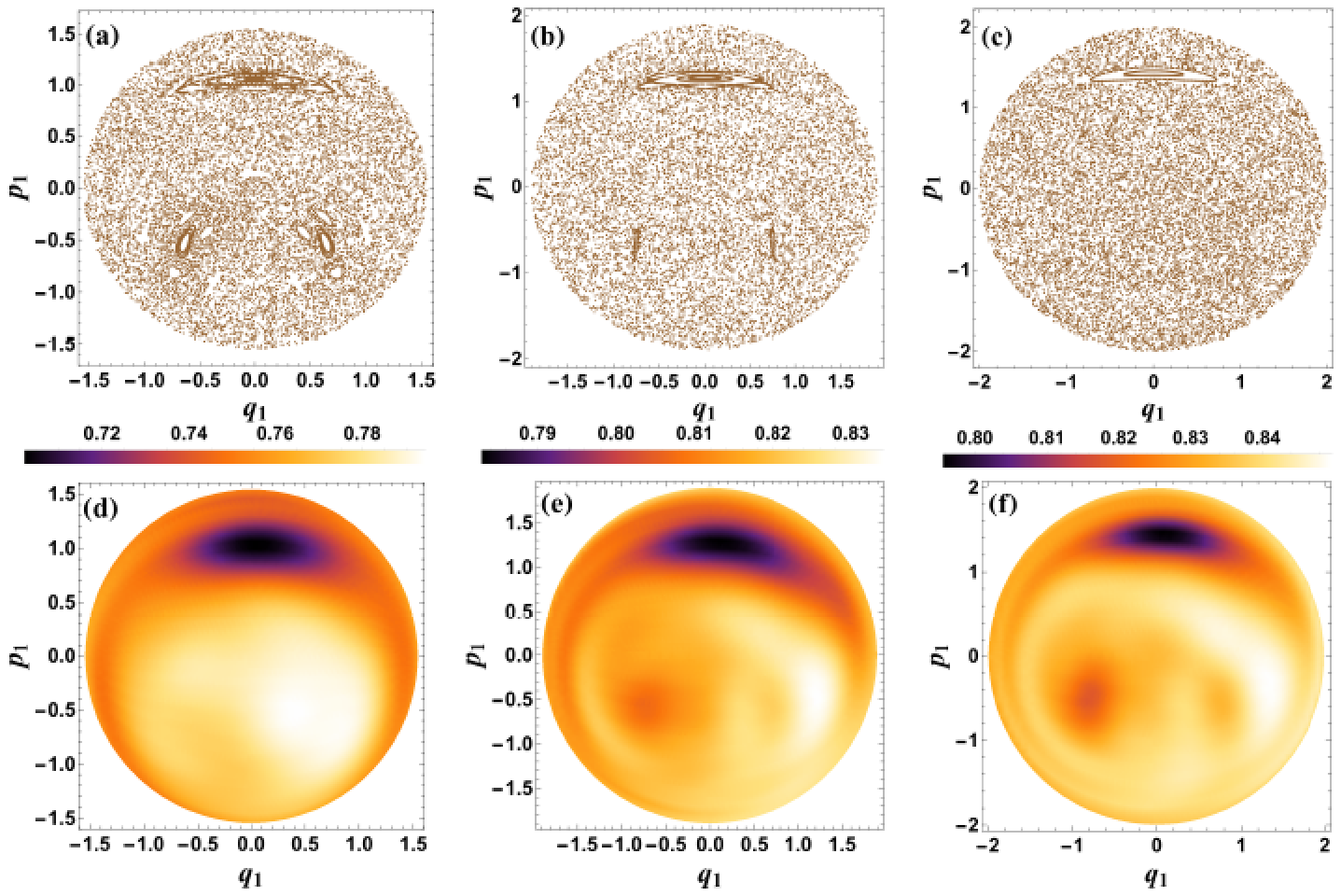}
\caption{The correspondence between classical dynamics and quantum entanglement entropy for the single photon Dicke model with the fixed particle number $N=10$ and coupling constant $g=0.5$. (a)-(c) correspond to the classical phase spaces for the one-photon Dicke model with the system energy $E=1$, $4$, $7$, respectively. (d)-(f) denote
the time-averaged entanglement entropy for the one-photon Dicke model with the system energy $E=1$, $4$, $7$, respectively. Here we set $\omega=\omega_{0}=1$. }\label{fd32}
\end{figure}
\begin{figure}
\includegraphics[width=11cm]{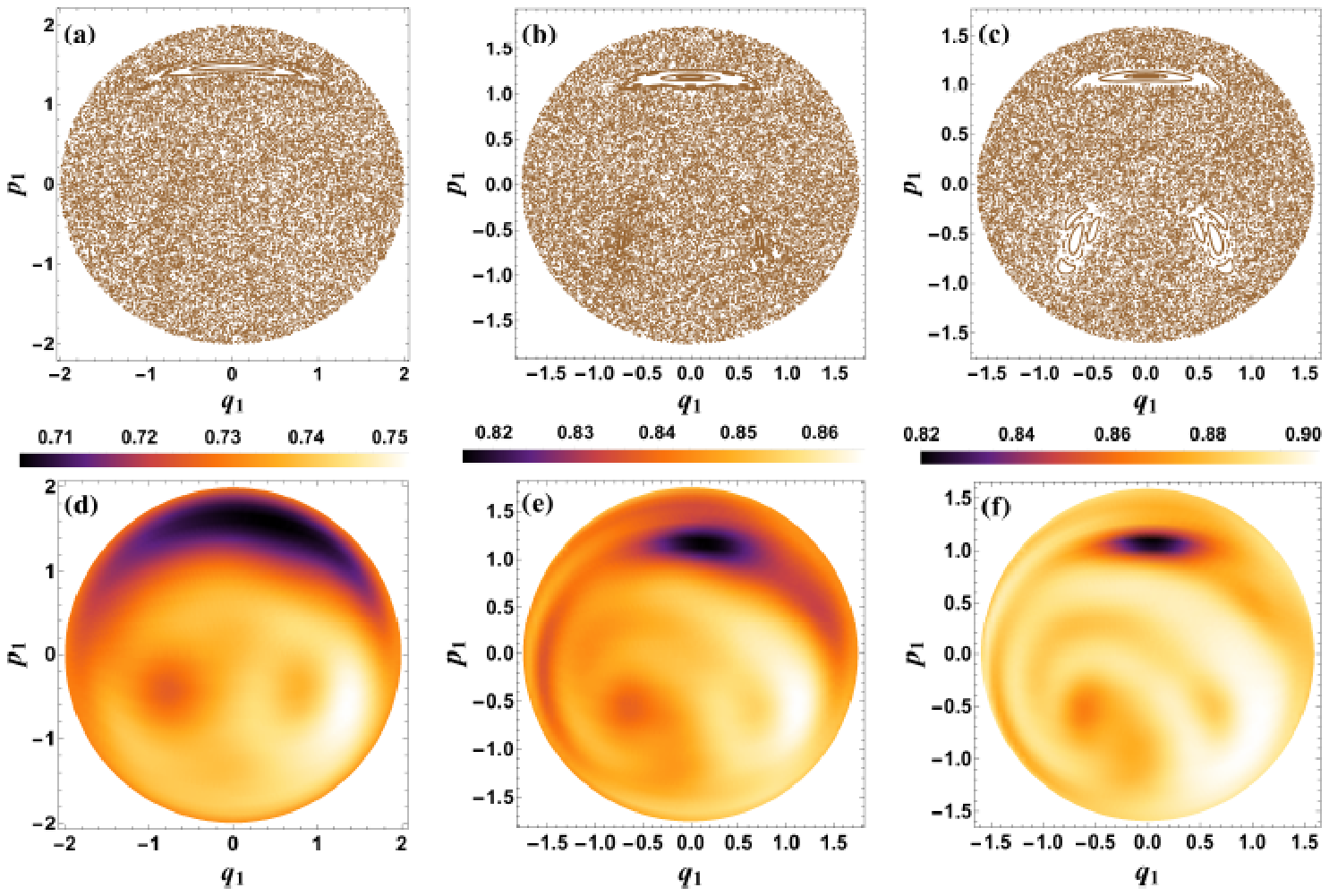}
\caption{The correspondence between classical dynamics and quantum entanglement entropy for the  one- photon Dicke model with the fixed system energy $E=4$ and coupling constant $g=0.5$. (a)-(c) correspond to the classical phase spaces for the single photon Dicke model with the particle number $N=5$, $15$, $30$, respectively. (d)-(f) denote the time-averaged entanglement entropy for the one-photon Dicke model with the particle number $N=5$, $15$, $30$, respectively. Here we set $\omega=\omega_{0}=1$.}\label{fd33}
\end{figure}
In order to make a comparison, in Fig.(\ref{fd3})-(\ref{fd33}), we plot the correspondence between classical dynamics and quantum entanglement entropy for the one-photon Dicke model. It is easy to find that the effects of the system energy $E$ and the particle number $N$ in the one-photon Dicke model are  similar to those in two-photon Dicke one. This also further supports that the increase of the system energy $E$ can improve the correspondence between classical dynamics and the time-averaged entanglement entropy.

\section{Husimi $Q$ distribution in the two-photon Dicke model}

Husimi $Q$ distribution is a quasiprobability distribution, which can provide a visualization of
highdimensional quantum states and demonstrates
the dynamical evolution of the quantum state with time. It is shown that
Husimi $Q$ distribution displays a rapid dispersion over the phase space as the initial coherent state is in the classically chaotic region. Thus, with Husimi $Q$ distribution, one can diagnose chaotic behavior in quantum system \cite{Husimi15}. For a coherent state, Husimi $Q$ function is defined as
\begin{eqnarray}
Q(q_{1},p_{1})&=&\frac{1}{\pi}\langle q_{1},p_{1}|\hat{\rho}_{1}|q_{1},p_{1}\rangle,
\end{eqnarray}
where $|q_{1},p_{1}\rangle$ is a coherent state and $\rho_{1}$ is the reduced-density matrix of the $1$st subsystem.
\begin{figure}[ht]
\includegraphics[width=13cm]{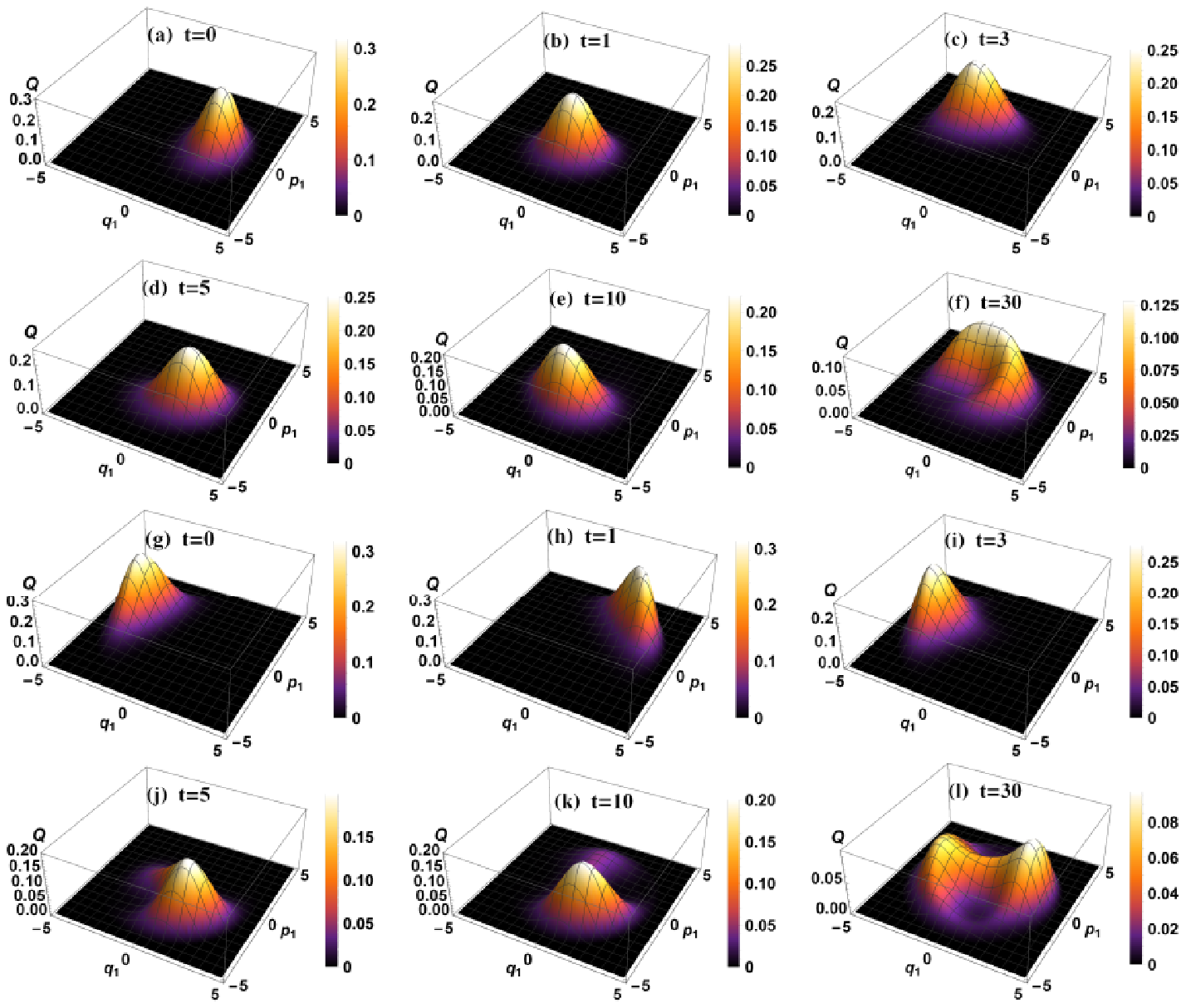}
\caption{ Change of Husimi $Q$ distribution with time for fixed coupling parameter $g=0.3$ and system energy $E=1$. (a)-(f) denote the case where the initial coherent state corresponds to the point $A_4$ in the regular region in Fig.\ref{f1}. (g)-(l) denote the case where the initial coherent state corresponds to the point $A_1$ in the chaotic region in Fig.\ref{f1}.}\label{f4}
\end{figure}
\begin{figure}
\includegraphics[width=13cm]{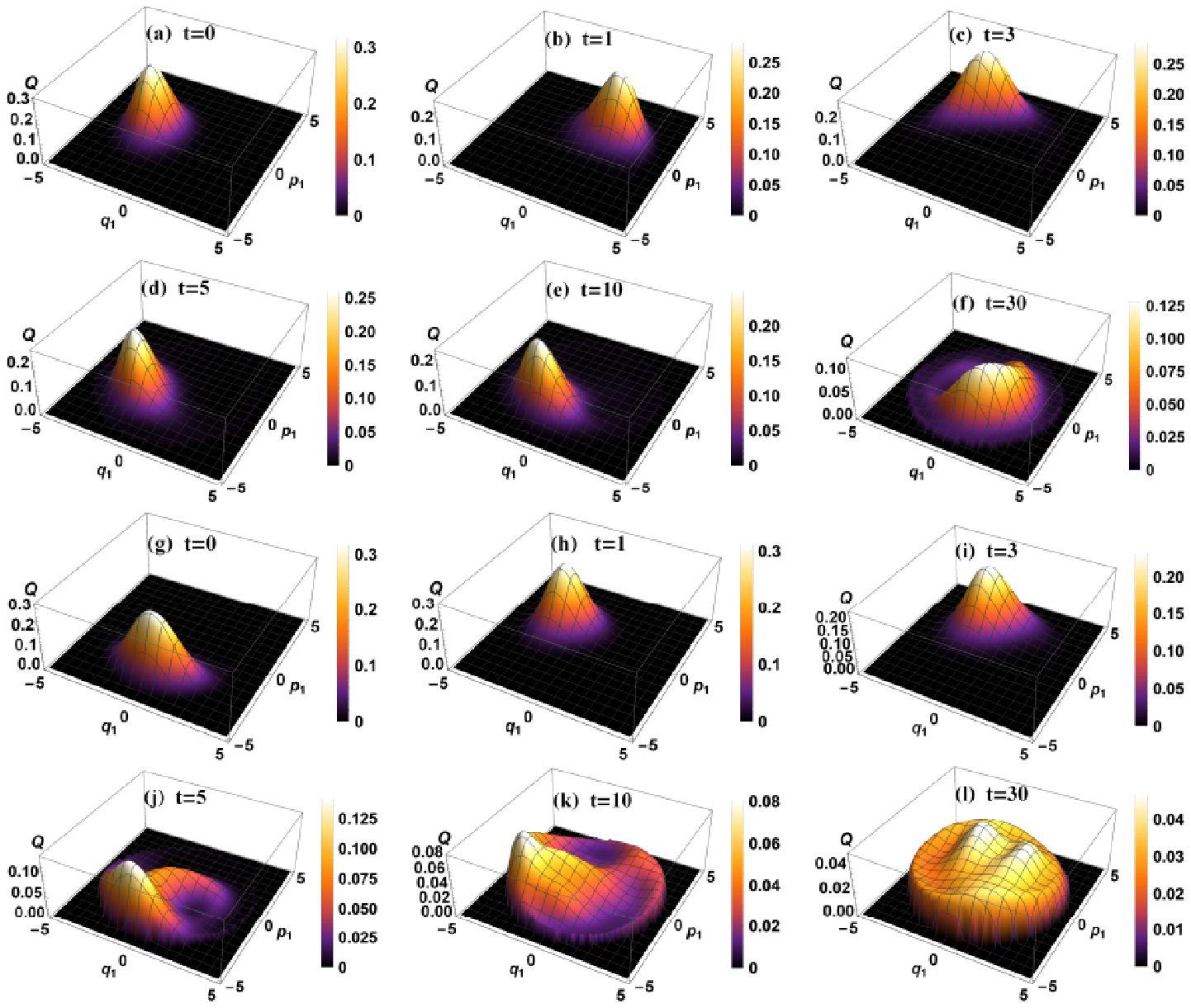}
\caption{Change of Husimi $Q$ distribution with time for fixed coupling parameter $g=0.3$ and system energy $E=10$. (a)-(f) denote the case where the initial coherent state corresponds to the point $C_1$ in the regular region in Fig.\ref{f1}. (g)-(l) denote the case where the initial coherent state corresponds to the point $C_3$ in the chaotic region in Fig.\ref{f1}.}\label{f5}
\end{figure}
\begin{figure}[ht]
\includegraphics[width=6cm]{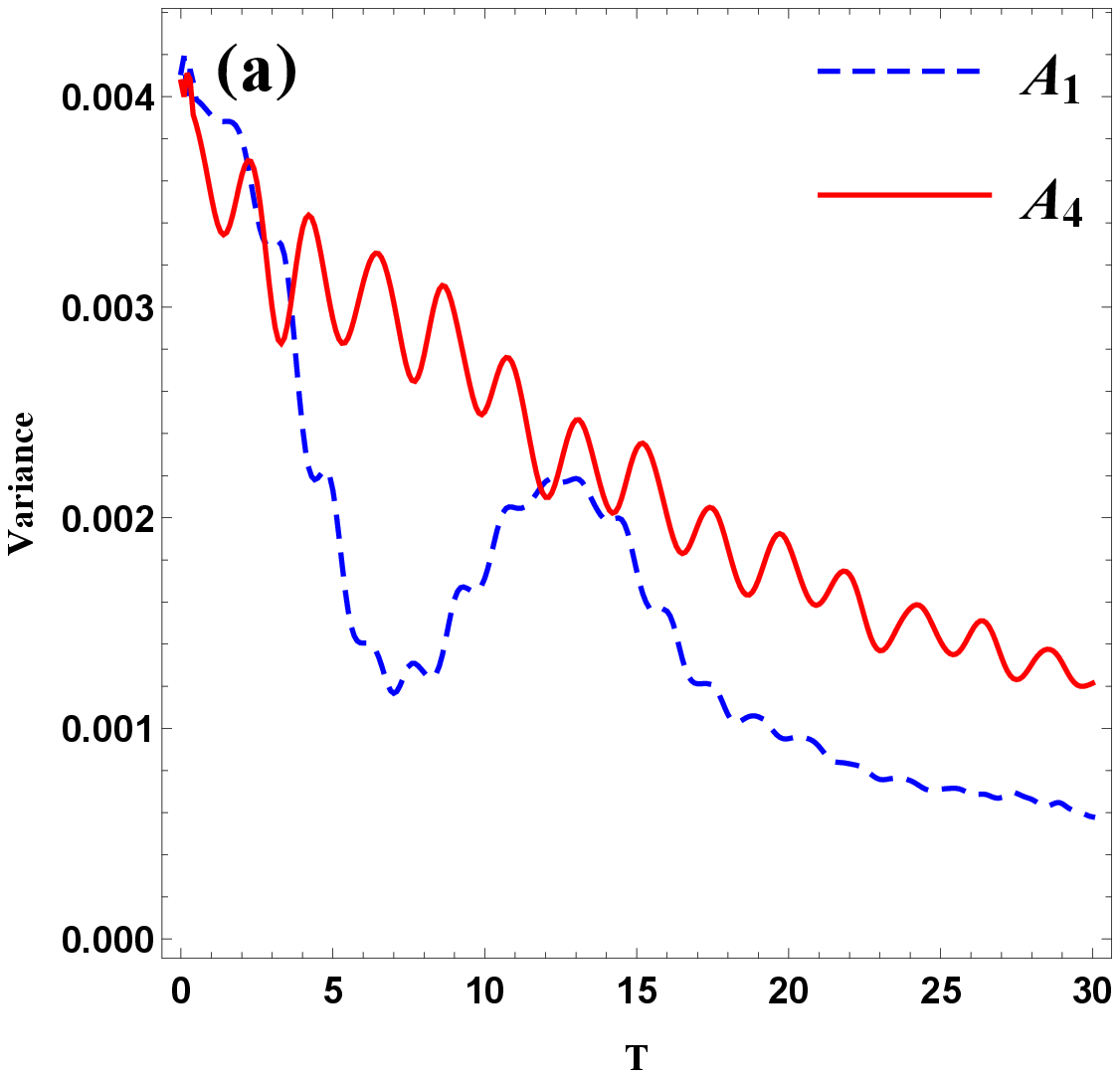}\;\;\;\;\;\;\;\;\includegraphics[width=6cm]{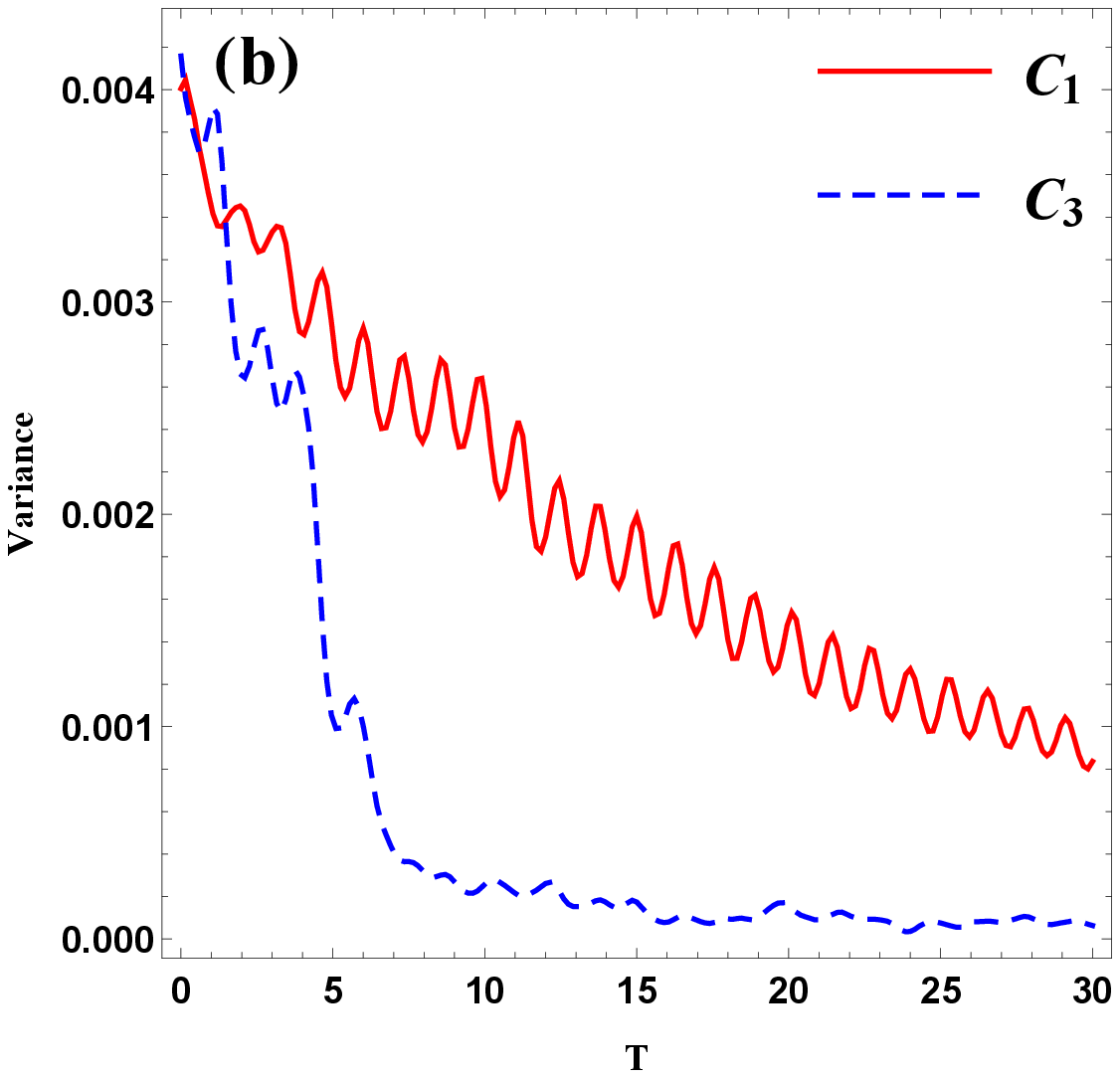}
\caption{Change of the variance in Husimi $Q$ distribution with time for fixed coupling parameter $g=0.3$ and $\omega=\omega_{0}/2=1$. In the left panel, the dashed blue and red lines  correspond to the initial coherent states located at the point $A_4$ in the regular region and the point $A_1$ in the chaotic region, respectively, in Fig.\ref{f1}(a) with the system $E=1$. In the right panel, the dashed blue and red lines  correspond to the initial coherent states located at the point $C_1$ in the regular region and the point $C_3$ in the chaotic region, respectively, in Fig.\ref{f1}(c) with the system $E=10$.}\label{f6}
\end{figure}
In Fig.\ref{f4} and Fig.\ref{f5}, we present the change of Husimi $Q$ distribution in phase space for the quantum system with energy $E=1$ and $E=10$, respectively. In Fig.\ref{f4}, (a)-(f) and (g)-(l) demonstrate the dynamical evolution of the coherent state with the initial state corresponding to the point $A_4$ in the regular region and the point $A_1$ in the chaotic region in Fig.\ref{f1} (a), respectively. Husimi $Q$ function owns almost the same dispersion rate in the phase space for these two different states, which indicates again that it is difficult to distinguish classical chaotic and regular behaviors in the two-photon Dicke model with the system energy $E=1$. However, for the case with the system energy $E=10 $ as shown in Fig.\ref{f5}, one can find Husimi $Q$ function for the initial state corresponded to point $C_3$ in the chaotic region disperses more quickly than that of the corresponding point $C_1$ in the regular region. Moreover, in Fig.\ref{f6}, we present the variance of Husimi $Q$ distribution for the quantum system with the same parameters as in Fig.\ref{f4} and Fig.\ref{f5}. It shows that the variance of Husimi $Q$ distribution for the initial state in the regular region decreases almost with the same rate in both cases $E=1$ and $E=10$. However, the variance in the chaotic region in the case $E=10$ decays more quickly than that in the case $E=1$. This means that the difference of the variance of Husimi $Q$ distribution between in the chaotic and regular regions is more distinct in the case with higher system energy. Therefore, the change of Husimi $Q$ distribution also supports that higher system energy has contributed to distinguish the chaotic and regular behavior in the quantum two-photon Dicke model.

Finally, we make a brief comparison of our results with other related studies. In the quantum kicked top model, it is found that the correspondence between the classical phase space and the distribution of time-averaged  entanglement entropy can be improved by increasing the particle number $N$ \cite{t14,enchao1,Ghose2019}, which is totally understandable since as $N$ (or $j$) tends infinite and the behavior of quantum system indeed converges to that in the classic limit. In the one-photon and two-photon Dicke models, we find that the high the system energy can improve the correspondence between the classical phase space and the distribution of time-averaged  entanglement entropy, which is not found elsewhere. This could be attributed to that at high system energy a quantum state would expect thermalization to a ``classical" high temperature state. Moreover, we also studied the effect of the particle number $N$ in the Dicke model on the correspondence between chaos and entanglement entropy, and found some similar results obtained in the quantum kicked top model. Therefore, our result indicate that the correspondence chaos and entanglement entropy can be improved by increasing both the particle number $N$ and the system energy in the Dicke models.

\section{Summary}

 We have studied entanglement entropy and Husimi $Q$ distribution in the two-photon Dicke model. It is shown that in the cases with higher system energy the increasing rate of linear entanglement  entropy becomes more rapidly for the initial states corresponded to these points in the classical chaotic region.  With the increase of the energy of system, the values of linear entanglement entropy of the points in these chaotic and regular regions become more distinguishable. Moreover,  there is an obvious improvement in the correspondence relationship between the distribution of time-averaged  entanglement entropy and the classical Poincar\'{e} section in the cases with higher system energy. Finally, we also present Husimi $Q$ distribution for a coherent state in the two-photon Dicke model with different system energies, and find that Husimi $Q$ distribution for the initial state corresponded to the point in the chaotic region in the higher energy system disperses more quickly than that in the lower energy system. These imply that higher system energy has contributed to distinguish the chaotic and regular behavior in the quantum two-photon Dicke model. Moreover,  our result indicate that in the Dicke model
 the correspondence between chaos and entanglement entropy can be improved by increasing both the particle number $N$ and the system energy $E$.  It would be of interest to study further whether there exist the inherent connection between these two different ways and whether there are other quantities which are susceptible to improve the correspondence chaos and entanglement entropy, which could be of benefit to
understand deeply the chaos in the quantum system.

\section{\bf Acknowledgments}
We would like to thank the anonymous referee for their useful comments and suggestions. This work was partially supported by the National Natural Science Foundation of China under
Grant No. 11875026, the Scientific Research
Fund of Hunan Provincial Education Department Grant
No. 17A124. J. Jing's work was partially supported by
the National Natural Science Foundation of China under
Grant No. 11875025.

\end{document}